\documentclass[twocolumn,showpacs,amsmath,amssymb,superscriptaddress]{revtex4-2}
\usepackage{dcolumn,braket,color,amsmath,footnote,hyperref,bm}
\usepackage{graphicx,bm,url,longtable}
\usepackage{tabularx,multirow,braket}
\usepackage{fancybox}
\usepackage{tikz}
\usetikzlibrary{matrix}

\begin{document}

\title{Microscopic description of octupole collective excitations near $N=56$ and $N=88$}

\author{K.~Nomura}
\email{knomura@phy.hr}
\author{L.~Lotina}
\author{T.~Nik\v si\'c}
\affiliation{Department of Physics, Faculty of Science, University of
Zagreb, HR-10000 Zagreb, Croatia}
\author{D.~Vretenar}
\affiliation{Department of Physics, Faculty of Science, University of
Zagreb, HR-10000 Zagreb, Croatia}
 \affiliation{ State Key Laboratory of Nuclear Physics and Technology,
 School of Physics, Peking University, Beijing 100871, China}

\date{\today}

\begin{abstract}
Octupole deformations and related collective excitations are analyzed 
using the framework of nuclear density 
 functional theory. Axially-symmetric quadrupole-octupole constrained
 self-consistent mean-field (SCMF) calculations with a choice of universal
 energy density functional and a pairing interaction are performed for
 Xe, Ba, and Ce isotopes from proton-rich to neutron-rich regions, and 
 neutron-rich Se, Kr, and Sr isotopes, 
%medium-heavy and heavy nuclei with the neutron and proton numbers $(N,Z)\approx (56,56)$, (88,56), and (56,34),  
%neutron-deficient nuclei with $N\approx Z\approx 56$ and neutron-rich
 %nuclei with $N\approx 56$ and $Z\approx 34$, 
in which enhanced octupole correlations are expected to
 occur. Low-energy positive- and  negative-parity spectra and
 transition strengths are computed by solving the quadrupole-octupole
 collective Hamiltonian, with the inertia parameters and
 collective potential determined by the 
 constrained SCMF calculations. 
Octupole-deformed equilibrium states are found in the potential
 energy surfaces of the Ba and Ce isotopes with 
 $N\approx 56$ and 88. The evolution of spectroscopic
 properties indicates enhanced octupole correlations in the regions 
 corresponding to $N\approx Z\approx 56$, $Z\approx 88$ and $Z\approx
 56$, and $N\approx 56$ and $Z\approx 34$. 
The average $\beta_{30}$ deformation parameter and its fluctuation exhibit 
 signatures of octupole shape phase transition around $N=56$ and 88. 
\end{abstract}

\maketitle

\section{Introduction}

The intrinsic shapes of most medium-heavy and heavy nuclei are characterized by reflection
symmetric, quadrupole deformations. Reflection-asymmetric, or octupole
deformations occurs in specific mass regions with the proton $Z$ and
neutron numbers $N$ near 34, 56, 88 and 134
\cite{butler1996,butler2016}. Octupole correlations determine  
the systematics of low-lying negative-parity states, which form
approximate alternating-parity doublets with the positive-parity 
ground-state bands, and the electric dipole and octupole
transition strengths. The exploration of stable octupole
deformations is a very active research field in both experimental and theoretical 
low-energy nuclear physics. In recent years, experiments with 
radioactive ion beams have identified octupole-deformed nuclei,
e.g., in light actinides ($^{220}$Rn, $^{222,224,228}$Ra, and
$^{228}$Th) \cite{gaffney2013,butler2020a,chishti2020}, and lanthanides
($^{144,146}$Ba) \cite{bucher2016,bucher2017}. 
Experimental studies of octupole deformations have also been reported in
lighter mass regions, e.g., the neutron-deficient nuclei with $N\approx Z\approx 56$ 
\cite{rugari1993,FAHLANDER1994,smith1998,SMITH2001,capponi2016}, and
neutron-rich nuclei with $N\approx 56$ and $Z\approx 34$ \cite{rzacaurban2000,gregor2017}. 

Theoretical analyses of octupole
deformations have used a variety of 
nuclear structure models, such as the self-consistent mean-field (SCMF) methods 
\cite{naza1984b,xia2017,agbemava2016,agbemava2017,cao2020}, 
the interacting boson model (IBM) \cite{engel1987,zamfir2001,zamfir2003,nomura2013oct,nomura2014}, the geometric
collective model \cite{bizzeti2013,bonatsos2015}, and the cluster
model \cite{shneidman2002,shneidman2003}. 
Most of these studies have been focused on
the regions corresponding to $Z\approx 88$ and 
$N\approx 134$, and $Z\approx 56$ and $N\approx 88$. 
However, octupole correlations in nuclei with particle numbers close to
$34$ and/or 56 have not been analyzed in much detail. 
A possible reason is that, especially because the $N\approx Z\approx 56$ nuclei
are close to the proton drip-line, experimental information is insufficient. 
Few exceptions are perhaps the Nilsson-Strutinsky calculation based on the 
Woods-Saxon potential in Refs.~\cite{naza1984b,SKALSKI1990}, the
constrained Hartree-Fock+BCS calculation with the Skyrme force
\cite{skyrme} of the 
light Xe and Ba isotopes in Ref.~\cite{heenen1994}, and the global
analysis of ground-state octupole deformation within the nuclear density
functional theory (DFT) in Ref.~\cite{cao2020}. However, 
in those studies calculations were carried out at the mean-field
level or only for restricted spectroscopic properties. 
Because of renewed experimental interest in octupole shapes in extended 
mass regions, it is meaningful to carry out a new theoretical
analysis of octupole deformations and related spectroscopy,
that also includes the lighter mass region with $N/Z\approx 34$ and 56. 

Nuclear DFT provides an accurate and economic 
microscopic approach to nuclear structure that enables systematic studies  \cite{bender2003,erler2011}. 
Both relativistic \cite{vretenar2005,niksic2011} and non-relativistic
 \cite{bender2003,robledo2019} energy density functionals (EDFs) have 
successfully been applied in the global description of the ground-state
 properties and collective excitations. The basic implementation is in terms 
 of SCMF calculations that produce energy 
 surfaces as functions of shape and/or pairing collective variables. 
To compute spectroscopic properties, the SCMF 
 framework must be extended to include dynamical correlations that
 arise from the restoration of broken symmetries and fluctuations around
 the mean-field minima. A straightforward approach is the generator coordinate
 method (GCM) \cite{RS} with symmetry
 projections and configuration mixing included. 
The GCM has been employed to study octupole
 correlations with axial quadrupole and octupole deformations as
 collective coordinates
 \cite{robledo2013,robledo2013,yao2015,bernard2016,lica2018,fu2018,rayner2020gcm}. 
In practical applications to medium-heavy and heavy nuclei, however, 
 the GCM is computationally challenging, especially as the
 number of nucleons or collective coordinates increases. 
Alternative approaches to GCM have thus been developed, such as the
 quadrupole-octupole collective Hamiltonian (QOCH)
 \cite{li2013,xia2017,xu2017} and the mapped $sdf$-IBM
 \cite{nomura2013oct,nomura2014}.

Based on the fully microscopic framework of nuclear DFT, here we
carry out a systematic analysis of 
octupole collective excitations in the mass $A\approx 90-150$ regions: 
Xe, Ba, and Ce isotopes extending from proton-rich ($N\approx
Z\approx 56$) to neutron-rich ($N\approx 88$ and $Z\approx 56$)
nuclei, and the neutron-rich Se, Kr, and Sr nuclei with $Z\approx 34$ and
$N\approx 56$. 
The starting point are axially-symmetric quadrupole-octupole
constrained SCMF calculations using the relativistic Hartree-Bogoliubov
model with the density-dependent point-coupling (DD-PC1) \cite{DDPC1}
EDF, and a separable pairing force \cite{tian2009}. 
The relevant excitation spectra and transition rates are computed
by solving the collective Schr\"odinger equation with the 
axially-symmetric quadrupole 
$\beta_{2}$ and octupole $\beta_{3}$ shape degrees of freedom. 
The constrained SCMF calculations completely determine the
moment of inertia, three mass parameters, and collective
potential of the QOCH. 
The diagonalization of the QOCH yields the positive- and
negative-parity excitation spectra, as well as the electric quadrupole,
octupole, and dipole transition rates. We note 
that a similar SCMF+QOCH spectroscopic calculation, based on the
PC-PK1 \cite{PCPK1} EDF, was performed for 
a large number of medium-heavy and heavy nuclei: from Rn to Fm, and from Xe to 
Gd isotopes \cite{xia2017}.  

Here we further mention recent EDF-based beyond SCMF calculations of the
octupole-related properties of nuclei. These include studies of, for
instances, the global systematics of octupole correlations in the ground
and excited states of 
virtually all even-even nuclei within the parity-projected GCM approach
using the Gogny EDFs \cite{robledo2011,robledo2015}, 
and the onset of octupole deformations and related spectroscopy in
neutron-rich Ba isotopes within the symmetry conserving configuration
mixing calculations with the 
Gogny-D1S EDF \cite{bernard2016,lica2018}, and within the multireference
covariant energy density functional theory \cite{fu2018} with
projections onto angular momentum, particle numbers, and parity. 

This paper is organised as follows. In Sec.~\ref{sec:theory} we briefly
review the formalism of the relativistic Hartree-Bogoliubov (RHB)+QOCH model. The SCMF 
$\beta_2-\beta_3$ potential energy surfaces are discussed in
Sec.~\ref{sec:pes}. 
In Sec.~\ref{sec:spectra} the systematics of spectroscopic properties,
including excitation energies of low-lying positive- and 
negative-parity states, and electromagnetic transition rates, are
compared to available experimental data. 
The results for the $N=56$ isotones are presented in
Sec.~\ref{sec:n56}. Signatures of octupole shape phase transitions
are examined in Sec.~\ref{sec:qpt}. Finally, a brief summary and
conclusion are given in Sec.~\ref{sec:summary}.

\section{Theoretical framework\label{sec:theory}}

\subsection{Relativistic Hartree-Bogoliubov calculation}

The first step of the analysis is a set of constrained SCMF 
calculations of potential energy surfaces (PESs), performed using 
the relativistic RHB method \cite{vretenar2005} 
with the DD-PC1 \cite{DDPC1} functional 
for the particle-hole channel, and a separable 
pairing force of finite range \cite{tian2009} 
in the particle-particle channel. 
The constraints imposed in the SCMF calculations 
are the expectation values of the 
axially-symmetric quadrupole $Q_{20}$ and octupole $Q_{30}$ moments:
\begin{align}
& \hat Q_{20} = 2z^2 - x^2 - y^2 \\
& \hat Q_{30} = 2z^3 - 3z(x^2 + y^2).
\end{align}
The corresponding 
quadrupole and octupole deformation parameters 
$\beta_{2}$ and $\beta_{3}$ are defined by the relations: 
\begin{align}
& \beta_2 = \frac{\sqrt{5\pi}}{3r_0^2 A^{5/3}} \braket{\hat Q_{20}} \\
& \beta_3 = \frac{\sqrt{7\pi}}{3r_0^3 A^{2}} \braket{\hat Q_{30}}, 
\end{align}
where $r_0 = 1.2$ fm. The calculations are performed in a harmonic oscillator (HO) 
basis with the number of oscillator shells $N_f=10$ for the region $Z\approx
34$ and $N\approx 56$. For heavier nuclei with $Z\approx 56$
and $N\geqslant 56$ a larger basis with
$N_F=12$ is used.

\subsection{Quadrupole-Octupole Collective Hamiltonian}

Collective states are described using an axially-symmetric QOCH, with
deformation-dependent parameters determined microscopically by the
constrained RHB calculation.  
The QOCH contains the vibrational and rotational kinetic terms, and 
the collective potential: 
\begin{align}
\label{eq:ham}
\hat H_\mathrm{coll} = {\mathcal T}_\mathrm{vib} + {\mathcal T}_\mathrm{rot} + V_\mathrm{coll},
\end{align}
where the vibrational  kinetic energy is parametrized by the mass parameters 
$B_{22}$, $B_{23}$, and $B_{33}$,  
\begin{align}
 {\mathcal T}_\mathrm{vib} = \frac{1}{2}B_{22}\dot\beta_2^2 +
 B_{23}\dot\beta_2\dot\beta_3 + \frac{1}{2}B_{33}\dot\beta_3^2, 
 \end{align}
 and the three moments of inertia $\mathcal{I}_k$ determine the rotational kinetic energy
 \begin{align}
{\mathcal T}_\mathrm{rot} = \frac{1}{2}\sum_{k=1}^{3}{\mathcal I}_k\omega_k^2. 
\end{align}
Finally, the collective potential $V_\mathrm{coll}$ includes zero-point energy (ZPE) 
corrections. 
After quantisation  the collective Hamiltonian reads: 
\begin{align}
 \hat H_\mathrm{coll} = &- \frac{\hbar^2}{2\sqrt{\omega\mathcal{I}}}
\Biggl[
\frac{\partial}{\partial\beta_2}\sqrt{\frac{\mathcal{I}}{\omega}}B_{33}\frac{\partial}{\partial\beta_2}
-
 \frac{\partial}{\partial\beta_2}\sqrt{\frac{\mathcal{I}}{\omega}}B_{23}\frac{\partial}{\partial\beta_3}
\nonumber \\
&-
 \frac{\partial}{\partial\beta_3}\sqrt{\frac{\mathcal{I}}{\omega}}B_{23}\frac{\partial}{\partial\beta_2}
+ \frac{\partial}{\partial\beta_3}\sqrt{\frac{\mathcal{I}}{\omega}}B_{22}\frac{\partial}{\partial\beta_3}
\Biggr] \nonumber \\
& + \frac{\hat J^2}{2\mathcal{I}} + V_\mathrm{coll}(\beta_2,\beta_3),
\end{align}
where $\omega = B_{22}B_{33} - B_{23}^2$. 
The mass parameters, moments of inertia, and collective potentials as functions of the collective coordinates  $(\beta_2, \beta_3)$, are specified by the deformation-constrained self-consistent RHB calculations for a specific choice of the nuclear energy density functional and pairing interaction. 
In the present version of the model, the mass parameters defined as the inverse of the mass tensor 
$B_{ij} ({\bf q}) = {\cal M} ^{-1}_{ij} ({\bf q})$,  
are calculated in the perturbative cranking approximation
 \begin{equation}
\label{eq:pmass}
{\cal M}^{C_p} = \hbar^2 {\it M}_{(1)}^{-1} {\it M}_{(3)} {\it M}_{(1)}^{-1}, 
\end{equation}
 where 
\begin{equation}
\label{eq:mmatrix}
\left[ {\it M}_{(k)} \right]_{ij} = \sum_{\mu\nu} 
    {\left\langle 0 \left| \hat{Q}_i \right| \mu\nu \right\rangle
     \left\langle \mu\nu \left| \hat{Q}_j \right| 0 \right\rangle
     \over (E_\mu + E_\nu)^k}\; . 
\end{equation}
$|\mu\nu\rangle$ are two-quasiparticle 
wave functions, and $E_\mu$ and $E_\nu$ the corresponding quasiparticle energies. $\hat{Q}_i$
denotes the multipole operators that correspond to the collective
degrees of freedom. 
The collective potential $V_{\mathrm{coll}}$ is obtained by subtracting the 
vibrational zero-point energy (ZPE) from the total RHB 
deformation energy
\begin{equation}
\label{eq:zpe}
E_{\rm ZPE} = {1\over4} {\rm Tr} \left[ {\it M}_{(2)}^{-1} {\it M}_{(1)} \right].
\end{equation}
The microscopic self-consistent solutions of the constrained RHB
equations, that is, the 
single-quasiparticle energies and wave functions on the entire energy surface as functions of the 
deformations, provide the microscopic input for the calculation of both the 
collective inertia and zero-point energy. 
The Inglis-Belyaev formula is used for the rotational moment of inertia.
From the diagonalization of the collective Hamiltonian (\ref{eq:ham})
one obtains the collective energy spectrum and eigenfunction. 
The eigenfunctions are expanded in terms of a complete set of
basis functions. For each value of the angular momentum $I$, the basis
is defined as: 
\begin{align}
 \ket{n_{2}n_{3}IMK} = (\omega\mathcal{I})^{-1/4}\phi_{n_2}(\beta_2)\phi_{n_3}(\beta_3)\ket{IMK},
\end{align}
where $\phi_{n_\lambda}$ denotes the one-dimensional HO
functions of $\beta_\lambda$. 
For positive (negative) parity states, $n_{3}$ and $I$ take even (odd) 
numbers. 
For axially-symmetric shapes, the intrinsic
projection of the total angular momentum $K=0$. 
The collective wave function is then expressed as:
\begin{align}
 \Psi_{\alpha}^{IM\pi}(\beta_2,\beta_3,\Omega)=\psi_{\alpha}^{I\pi}(\beta_2,\beta_{3})\ket{IM0},
\end{align}
with $\Omega$ representing three Euler angles. 
The corresponding probability density distribution is defined as: 
\begin{align}
\label{eq:rho}
 \rho^{I\pi}_{\alpha}(\beta_2,\beta_3)=\sqrt{\omega\mathcal{I}}|\psi_{\alpha}^{I\pi}(\beta_2,\beta_{3})|^2,
\end{align}
with the normalization
\begin{align}
 \int\rho_{\alpha}^{I\pi}(\beta_2,\beta_3)d\beta_2d\beta_3 = 1.
\end{align}

The reduced transition probabilities $B(E\lambda)$ are calculated from the
relation: 
\begin{align}
 B(E\lambda;I_i\to I_f) = 
&(I_{i}0\lambda 0|I_{f}0)^2 \nonumber \\
&\times\Biggl|\int
 d\beta_{2}d\beta_{3}\sqrt{\omega\mathcal{I}}
\psi_i
{\mathcal{M}}_{E\lambda}(\beta_2,\beta_3)
\psi_{f}^\ast
\Biggr|^{2},
\end{align}
where ${\mathcal{M}}_{E\lambda}(\beta_2,\beta_3)$ is the electric moment
of order $\lambda$, and the factor in parentheses on the right-hand side
of the above expression is the Clebsch-Gordan coefficient. The electric
moment is calculated as
$\braket{\Phi(\beta_2,\beta_3)|\hat{\mathcal{M}}_{E\lambda}|\Phi(\beta_2,\beta_3)}$, with $\Phi(\beta_2,\beta_3)$ representing the wave functions obtained
from the RHB calculations. 
The corresponding operators $\hat{\mathcal{M}}_{E\lambda}$ for dipole,
quadrupole, and octupole transitions read: 
\begin{align}
& D_1 = \sqrt{\frac{3}{4\pi}}e\Bigl(\frac{N}{A}z_p -
 \frac{Z}{A}z_n\Bigr) \\
& Q_2^p = \sqrt{\frac{5}{16\pi}}e(2z_p^2 - x_p^2 - y_p^2) \\
& Q_3^p = \sqrt{\frac{7}{16\pi}}e(2z_p^3 - 3z_p (x_p^2 + y_p^2)), 
\end{align}
respectively, where bare electric charge $e$ is used, and this means no
effective charges need to be introduced to calculate electromagnetic
transition rates.

%-----------------------------------------------------------------------
%
%	PESs for Xe-Ba-Ce
%
%-----------------------------------------------------------------------
\begin{figure*}[htb!]
\begin{center}
\includegraphics[width=0.75\linewidth]{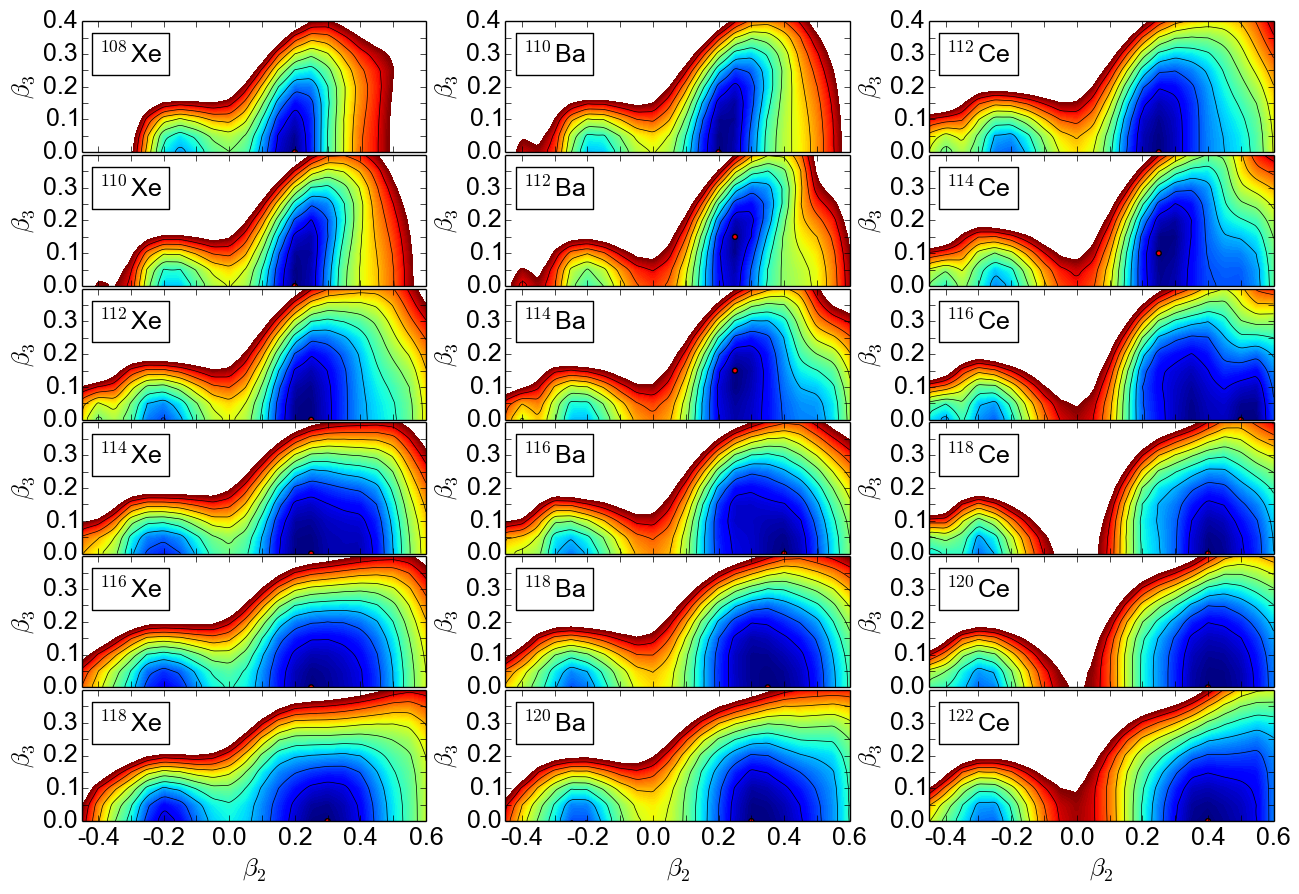}
\caption{SCMF $(\beta_2,\beta_3)$ PESs for 
 $^{108-118}$Xe, $^{110-120}$Ba
 and
 $^{112-122}$Ce. Global minima are identified by
 the red dots. Contour joints points on the surface with the same
 energy, and the difference between neighbouring contours is 1 MeV.} 
\label{fig:pes-prich}
\end{center}
\end{figure*}

%-----------------------------------------------------------------------
%
%	PESs for Xe-Ba-Ce
%
%-----------------------------------------------------------------------
\begin{figure*}[htb!]
\begin{center}
\includegraphics[width=0.75\linewidth]{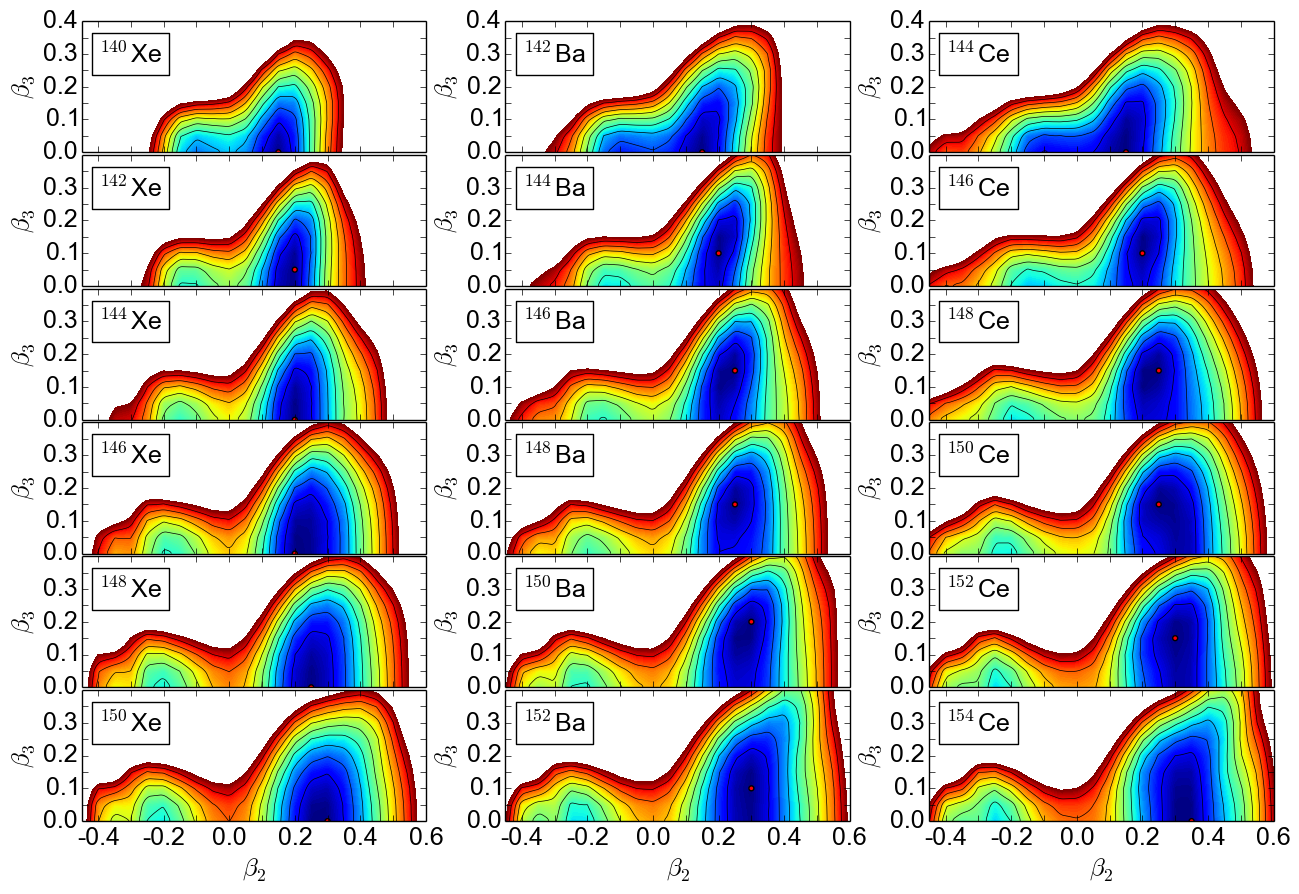} 
\caption{Same as in the caption to Fig.~\ref{fig:pes-prich} but for
 $^{140-150}$Xe, $^{142-152}$Ba, and $^{144-154}$Ce.}
\label{fig:pes-nrich}
\end{center}
\end{figure*}

%-----------------------------------------------------------------------
%
%	PESs fo Se-Kr-Sr
%
%-----------------------------------------------------------------------
\begin{figure*}[htb!]
\begin{center}
\includegraphics[width=0.75\linewidth]{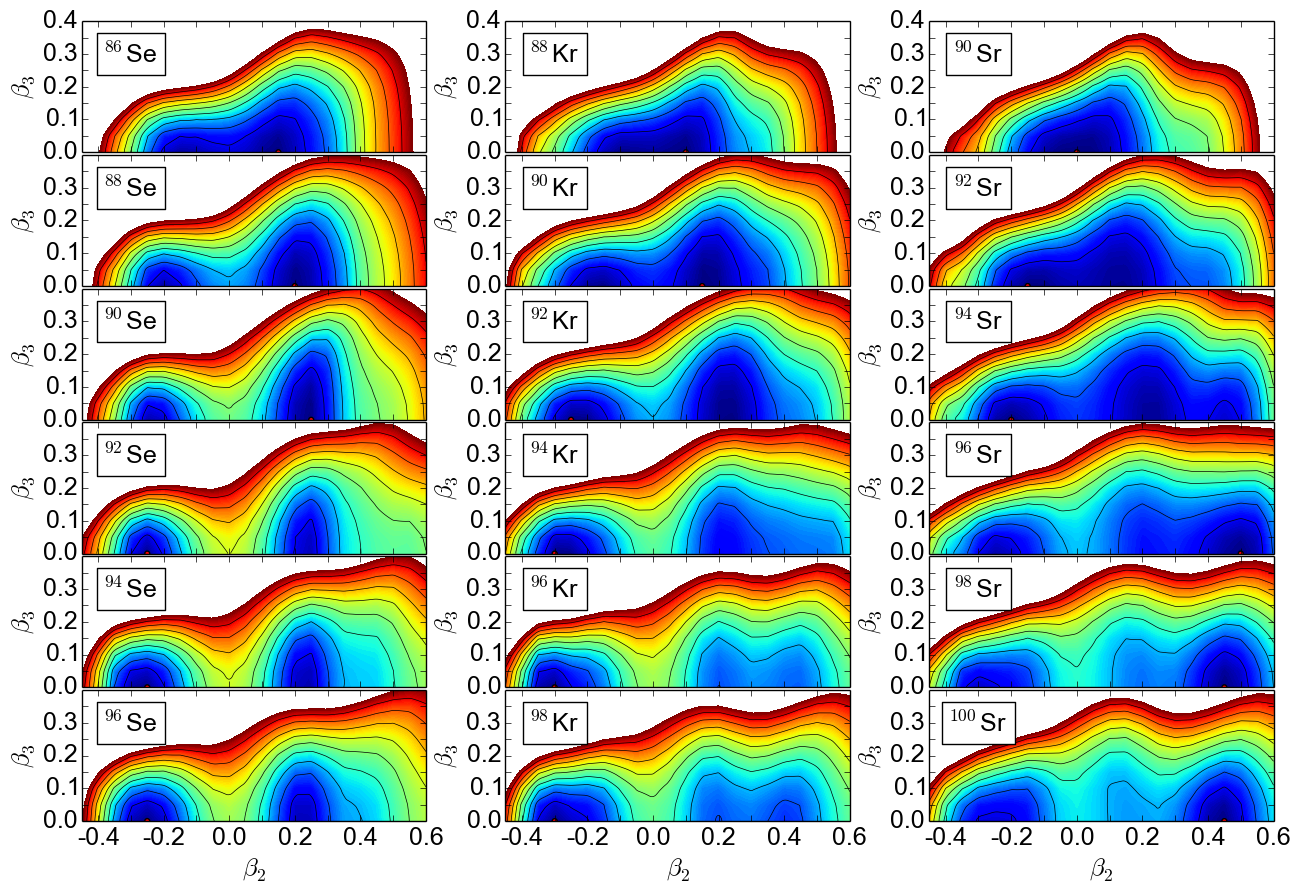} 
\caption{Same as in the caption to Fig.~\ref{fig:pes-prich} but for the
 neutron-rich nuclei $^{86-96}$Se, $^{88-98}$Kr, and
 $^{90-100}$Sr.}
\label{fig:pes-kr}
\end{center}
\end{figure*}

%-----------------------------------------------------------------------
%
%	PESs for N=56 isotones
%
%-----------------------------------------------------------------------
\begin{figure}[htb!]
\begin{center}
\includegraphics[width=\linewidth]{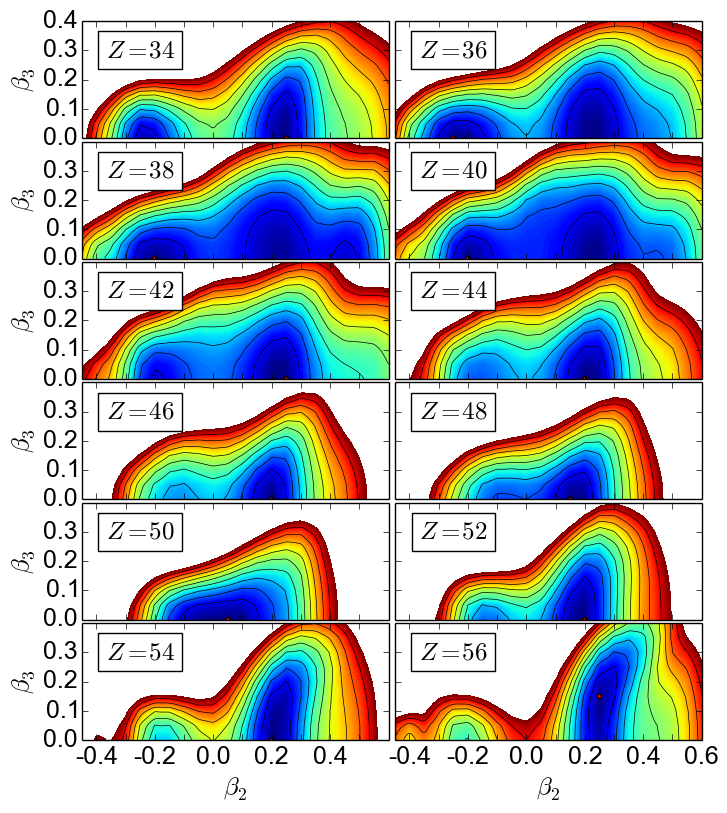} 
\caption{Same as in the caption to Fig.~\ref{fig:pes-prich} but for the
 $N=56$ isotones from $^{90}$Se ($Z=34$) to $^{112}$Ba ($Z=56$).} 
\label{fig:pes-n56}
\end{center}
\end{figure}

\section{SCMF results\label{sec:pes}}

\subsection{Neutron-deficient $Z\approx 56$ nuclei}

The axially-symmetric $(\beta_2,\beta_3)$ PESs for neutron-deficient
nuclei: $^{108-118}$Xe, $^{110-120}$Ba and $^{112-122}$Ce are depicted in
Fig.~\ref{fig:pes-prich}. 
Already at the most neutron-deficient isotopes with $N=54$, the
potential is considerably soft 
in $\beta_3$ deformation, even though the minimum is on the $\beta_3=0$ axis. 
Octupole-deformed equilibrium states with $\beta_3\neq 0$ occur in the
$N\approx Z$ nuclei $^{112,114}$Ba and $^{114}$Ce. There is no stable octupole
deformed minimum for the neighbouring Xe isotopes, but the potential
exhibits a narrow valley on the prolate side ($\beta_2 > 0$) that is
soft over a range of $\beta_3$ values. 
Previous mean-field calculations have also suggested there are a few $N\approx
Z\approx 56$ nuclei that exhibit octupole-deformed equilibrium states on
the potential energy surfaces
\cite{SKALSKI1990,heenen1994,agbemava2016,cao2020}. 
For $N>64$, not shown in the figure, the potential
becomes rather softer in the $\beta_{2}$ direction and the prolate
deformation becomes larger around the middle of the major shell $N=66$, but no
octupole minima are found on the corresponding SCMF PESs. When 
approaching the neutron shell closure at $N=82$, nearly spherical global
minima are obtained with both the $\beta_{2}$ and $\beta_{3}$
deformations converging to zero.

\subsection{Neutron-rich $Z\approx 56$ nuclei}

Figure \ref{fig:pes-nrich} displays the $(\beta_2,\beta_3)$ PESs for the isotopes $^{140-150}$Xe,
$^{142-152}$Ba, and $^{144-154}$Ce beyond the $N=82$ neutron
shell closure. These 
neutron-rich isotopes are close to the empirical octupole magic
number $N=88$, and more extensive experimental and theoretical
studies have been reported in this region compared to the
neutron-deficient one with $N\approx Z\approx 56$. 
In all three isotopic chains the potential surfaces shown in
Fig.~\ref{fig:pes-nrich} are more rigid in $\beta_2$, and 
pronounced octupole correlations are predicted. 
In particular, a number of neutron-rich Ba and 
Ce nuclei exhibit octupole-deformed  global minima with non-zero value
of $\beta_3$, that is, the isotopes $^{144-152}$Ba and
$^{146-152}$Ce. 
The most pronounced octupole global minimum is found in nuclei
with $N\approx 88$, in agreement with experimental findings.
The $\beta_{2}-\beta_{3}$ PESs obtained in the present analysis for the
neutron-rich lanthanides are also consistent with many of 
the recent SCMF calculations using both relativistic
\cite{nomura2014,agbemava2016,xia2017,fu2018} and non-relativistic EDFs
\cite{bernard2016,lica2018,cao2020}.

\subsection{$Z\approx 34$ nuclei around $N=56$}

We will also explore another mass region in which octupole
correlations could develop. 
In Fig.~\ref{fig:pes-kr} we plot the SCMF $\beta_2-\beta_3$ PESs for the
neutron-rich nuclei $^{86-96}$Se, $^{88-98}$Kr, and $^{90-100}$Sr, 
close to the proton $Z=34$ and neutron $N=56$ octupole magic numbers. 
Even though octupole correlations are empirically expected to occur at
proton number $Z=34$, the PESs in the figure do not exhibit octupole
global minima for these nuclei. In general, the
$(\beta_2,\beta_3)$ PESs for the $Z\approx 34$ neutron-rich
nuclei appear rather soft in $\beta_2$ deformation. Taking as example
$^{96,98}$Kr,  
one notices two shallow local minima on the prolate side. 
For many nuclei in this region a number of both microscopic and
empirical studies point to the presence of shape coexistence and/or
$\gamma$-soft shapes. The present calculation is restricted to 
only axially-symmetric shapes and, thus, a more realistic 
analysis should take into account the triaxial degrees of
freedom.

\subsection{$N=56$ isotones}

To analyze the evolution of the empirical $N=56$ octupole magic number, 
we have performed constrained SCMF calculations along the isotonic 
chain $N=56$. Figure~\ref{fig:pes-n56} displays the
resulting $\beta_2-\beta_3$ PESs for the $N=56$ even-even isotones from
$Z=34$ ($^{90}$Se) to $Z=56$ ($^{112}$Ba). 
In the $(\beta_2,\beta_3)$ PESs, neither an octupole deformed equilibrium state 
nor octupole-soft potential is observed below the proton magic number
$Z=50$. For $N=56$ isotones beyond Sn, however, the
potentials start to become more rigid in $\beta_2$ and softer in
$\beta_3$. Among the $N=56$ isotones depicted in the figure, the most
pronounced octupole minimum is obtained for the nucleus $^{112}$Ba with
$N=Z=56$.

%-----------------------------------------------------------------------
%
%	Level Scheme for Ba-144, Xe-112
%
%-----------------------------------------------------------------------
\begin{figure}[htb!]
\begin{center}
\includegraphics[width=\linewidth]{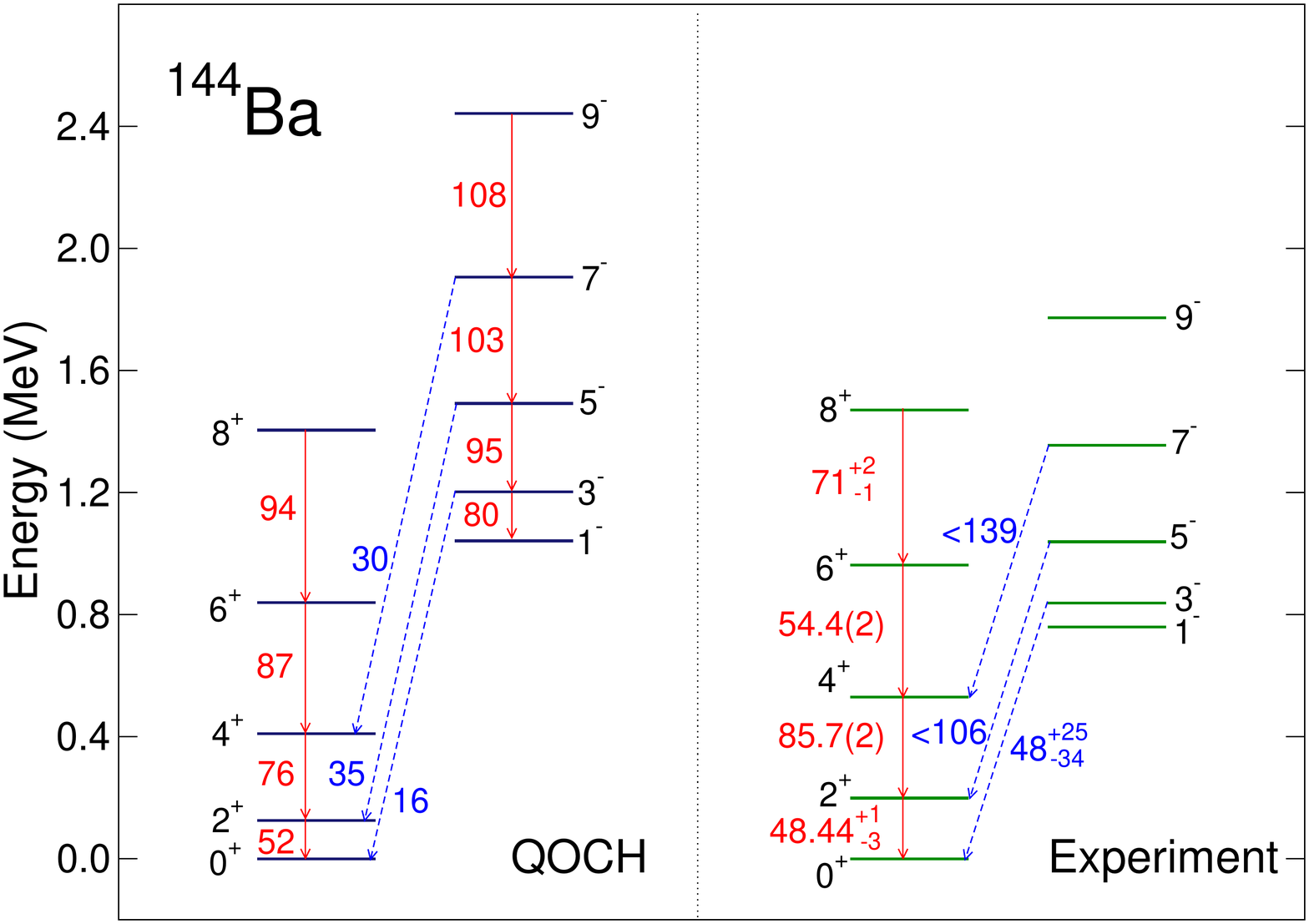} \\
\includegraphics[width=\linewidth]{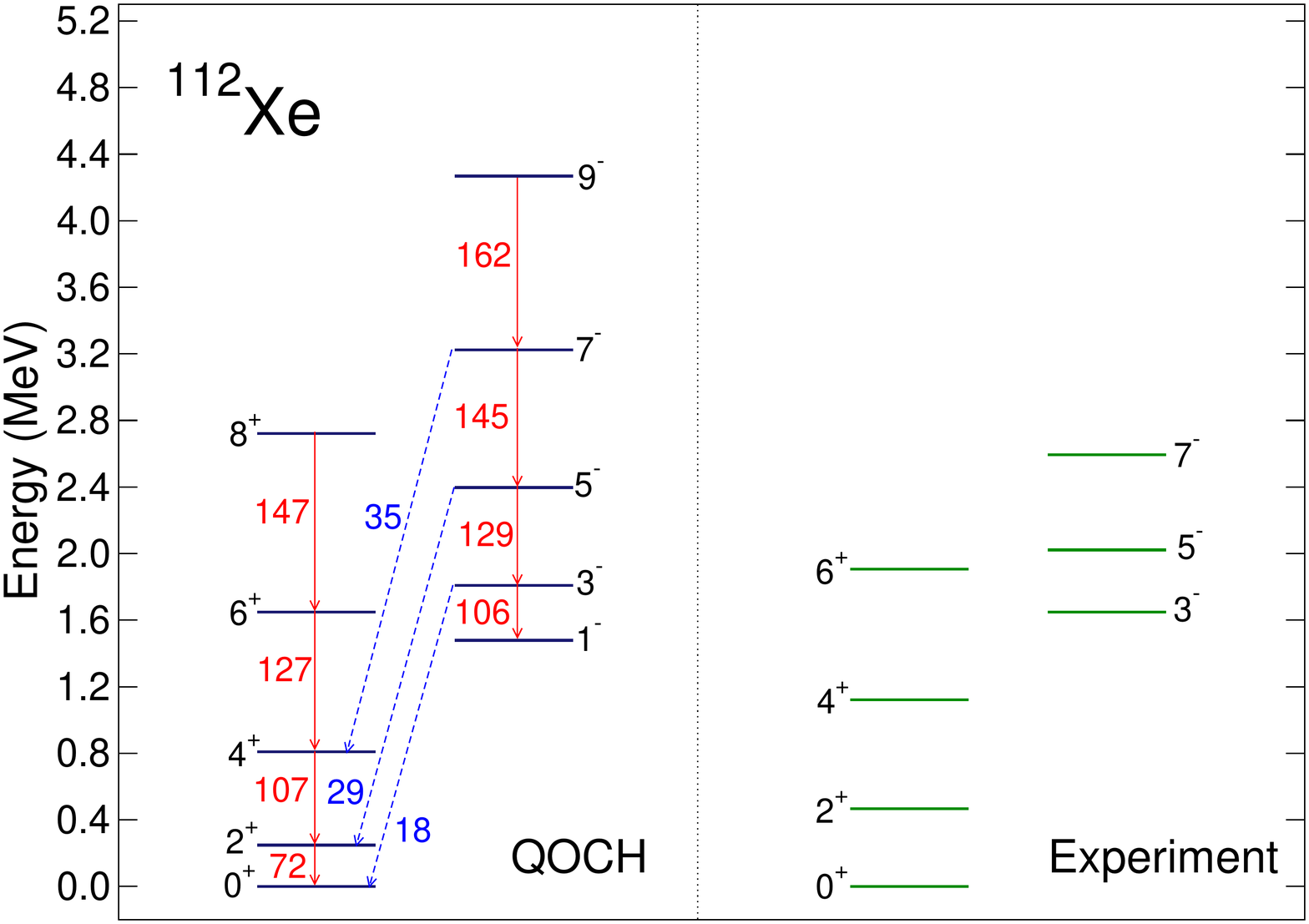} 
\caption{Comparison of the QOCH and experimental
 low-energy excitation spectra for the positive- and negative-parity
 yrast states of $^{144}$Ba and $^{112}$Xe. Solid and
 dashed arrows denote E2 and E3 transitions, respectively, and the 
 corresponding $B(E2)$ and $B(E3)$ values are given in Weisskopf
 units. Experimental results are from Refs.~\cite{bucher2016,data}.}
\label{fig:level-baxe}
\end{center}
\end{figure}

%-----------------------------------------------------------------------
%
%	Collective w.f. Ba-144, Xe-112
%
%-----------------------------------------------------------------------
\begin{figure}[htb!]
\begin{center}
 \includegraphics[width=\linewidth]{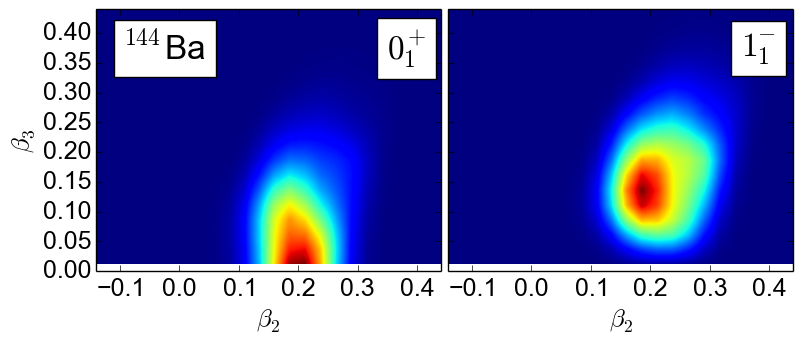} \\
\includegraphics[width=\linewidth]{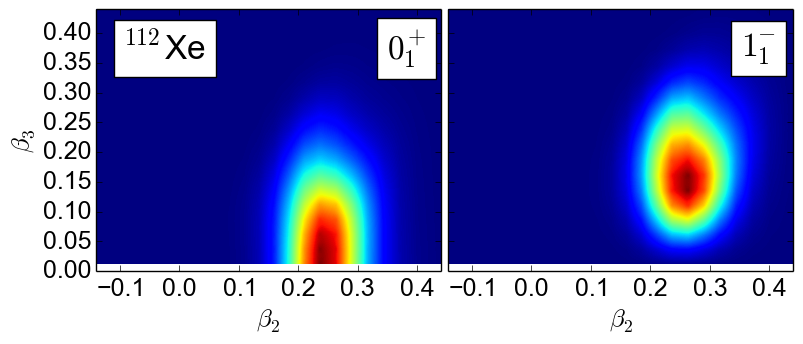}
\caption{Probability density distributions for the lowest
 positive-parity ($0^+_1$) and negative-parity ($1^-_1$) states of 
 $^{144}$Ba (upper) and $^{112}$Xe (lower) in the $\beta_2-\beta_3$ plane.} 
\label{fig:cwf-baxe}
\end{center}
\end{figure}

%-----------------------------------------------------------------------
%
%	SPECTRA -- Xe-Ba-Ce
%
%-----------------------------------------------------------------------
\begin{figure*}[htb!]
\begin{center}
\includegraphics[width=0.75\linewidth]{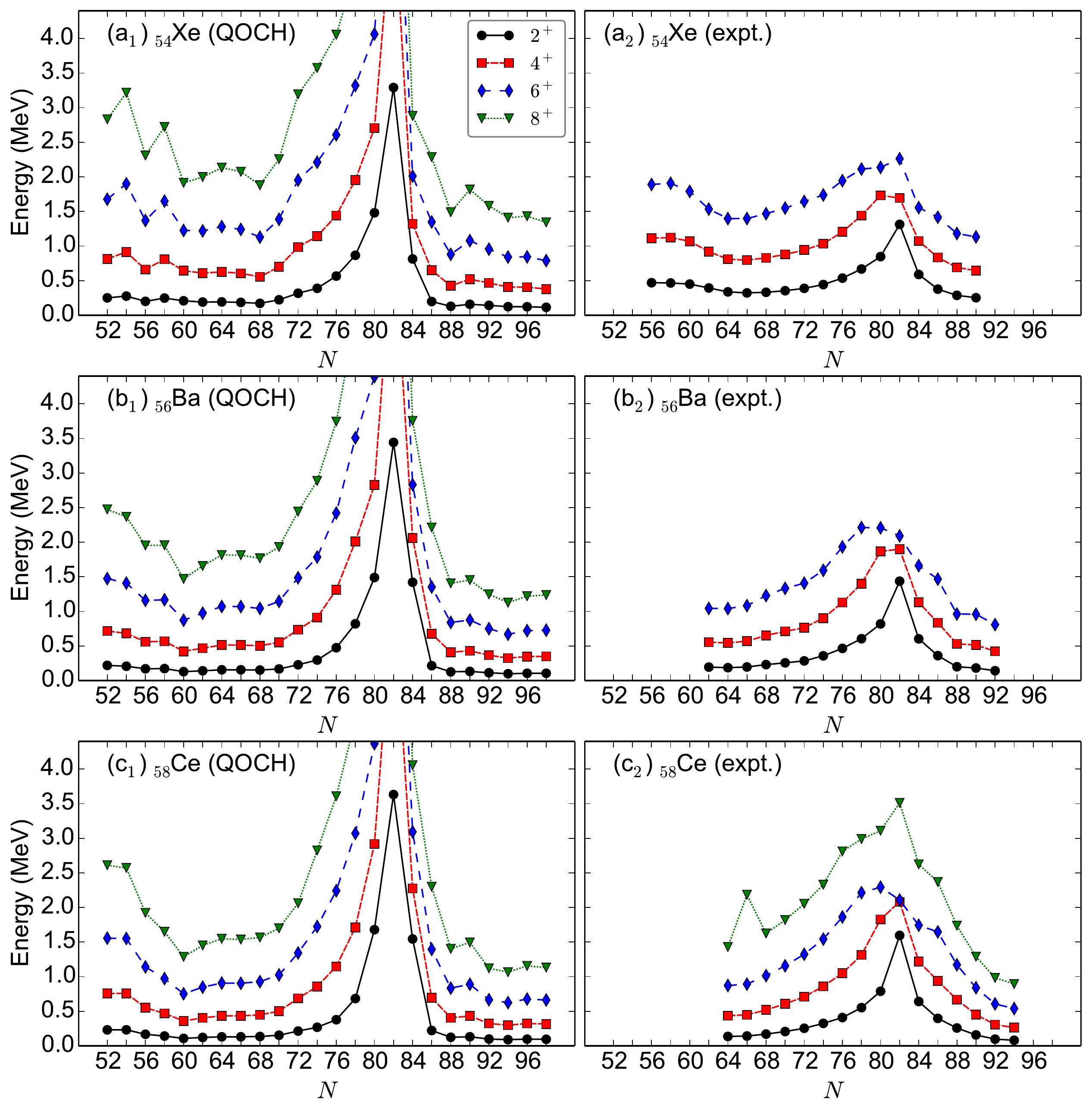} 
\caption{Evolution of QOCH excitation spectra for the 
 positive-parity yrast states along the chains of
 Xe, Ba, and Ce isotopes. Experimental values are from the ENSDF database \cite{data}.}
\label{fig:xe-level-pos}
\end{center}
\end{figure*}

%-----------------------------------------------------------------------
%
%	SPECTRA -- XE, BA, CE
%
%-----------------------------------------------------------------------
\begin{figure*}[htb!]
\begin{center}
\includegraphics[width=0.75\linewidth]{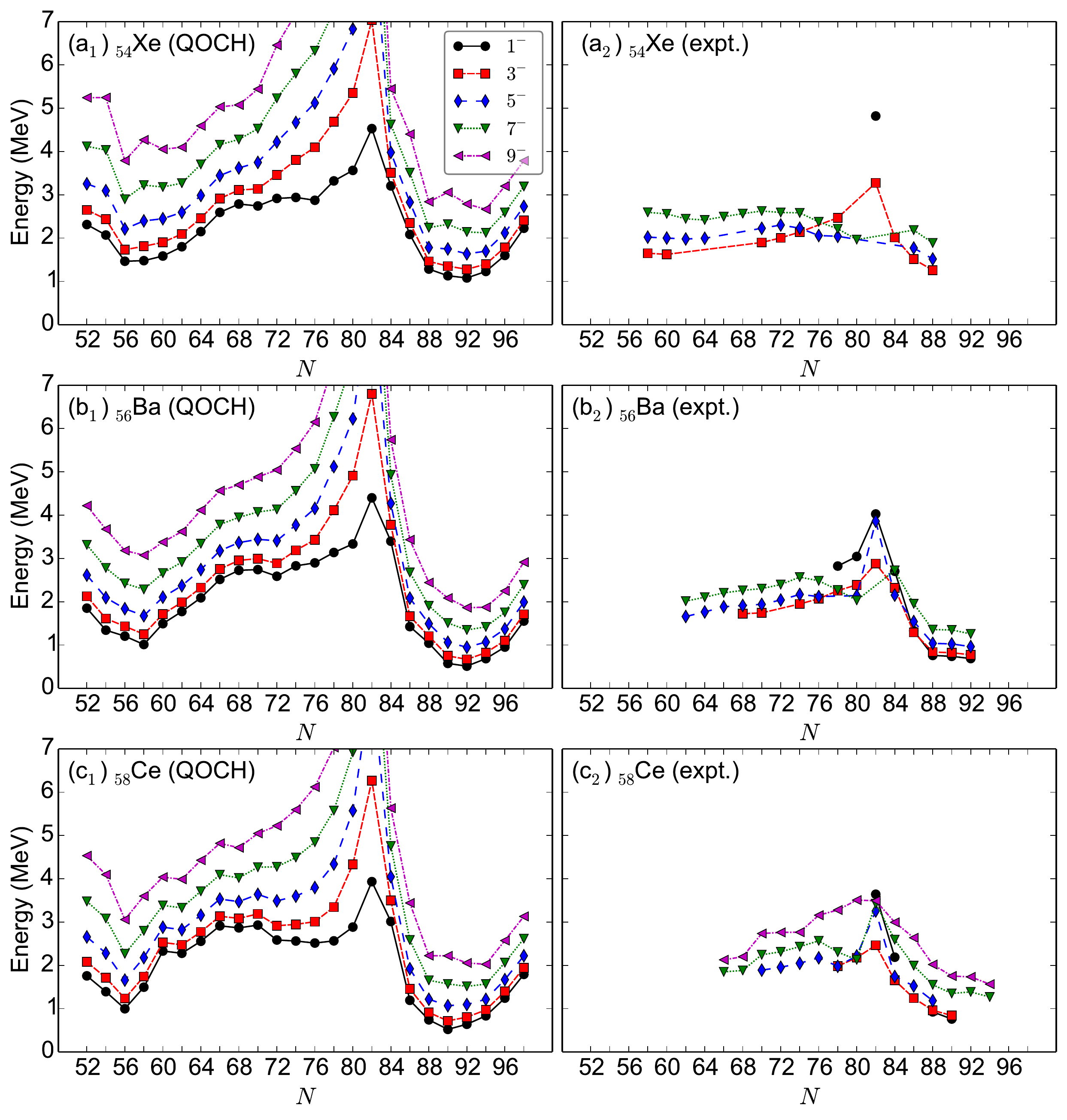} 
\caption{Same as in the caption to
 Fig.~\ref{fig:xe-level-pos} but for the negative-parity states.} 
\label{fig:xe-level-neg}
\end{center}
\end{figure*}

%-----------------------------------------------------------------------
%
%	SPECTRA -- Sr, Kr, Se, positive parity
%
%-----------------------------------------------------------------------
\begin{figure}[htb!]
\begin{center}
\includegraphics[width=\linewidth]{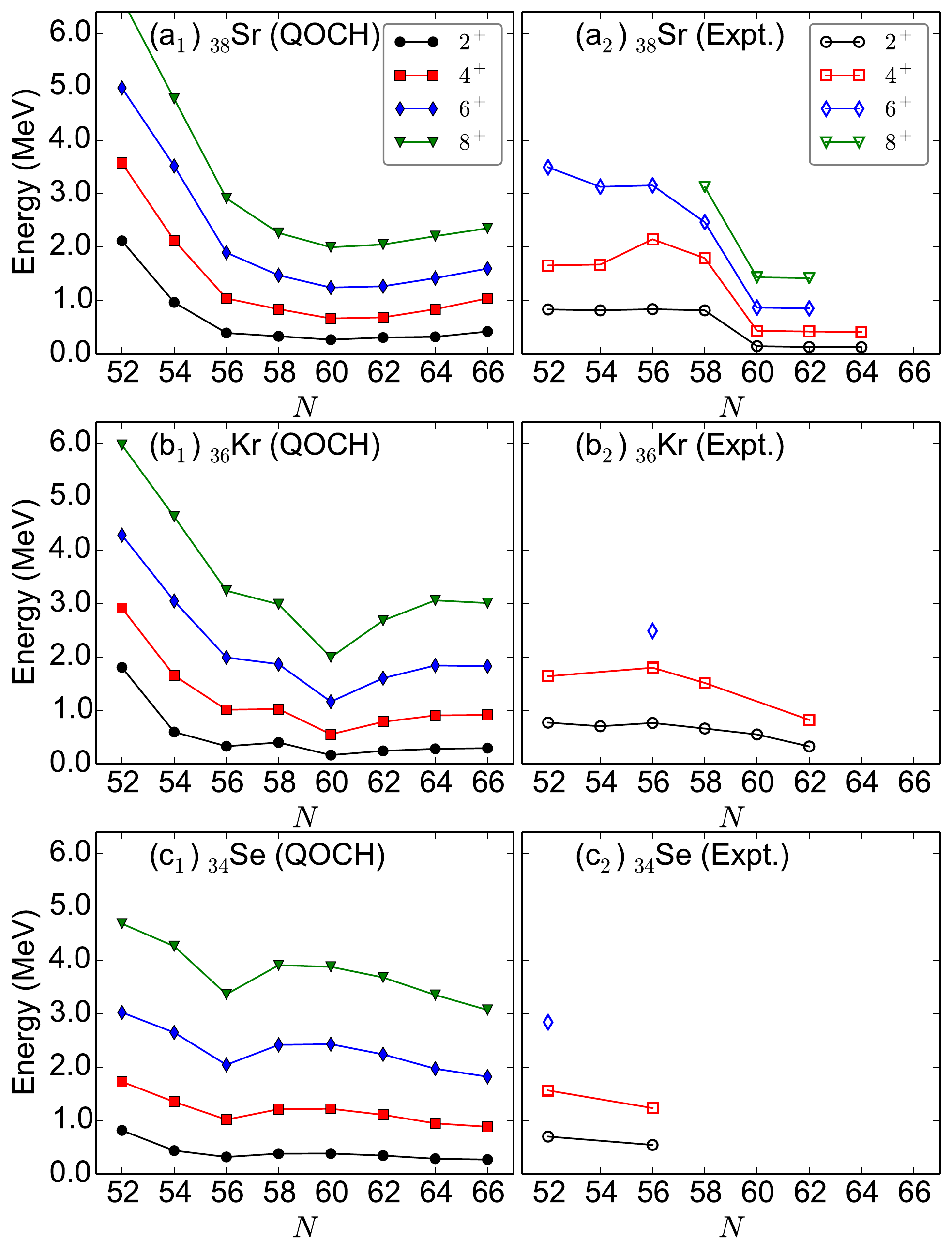} 
\caption{Same as in the caption to
 Fig.~\ref{fig:xe-level-pos} but for the positive-parity states in Se, Kr, and Sr isotopes. 
} 
\label{fig:level-kr-pos}
\end{center}
\end{figure}

%-----------------------------------------------------------------------
%
%	SPECTRA -- Sr, Kr, Se, negative parity
%
%-----------------------------------------------------------------------
\begin{figure}[htb!]
\begin{center}
\includegraphics[width=\linewidth]{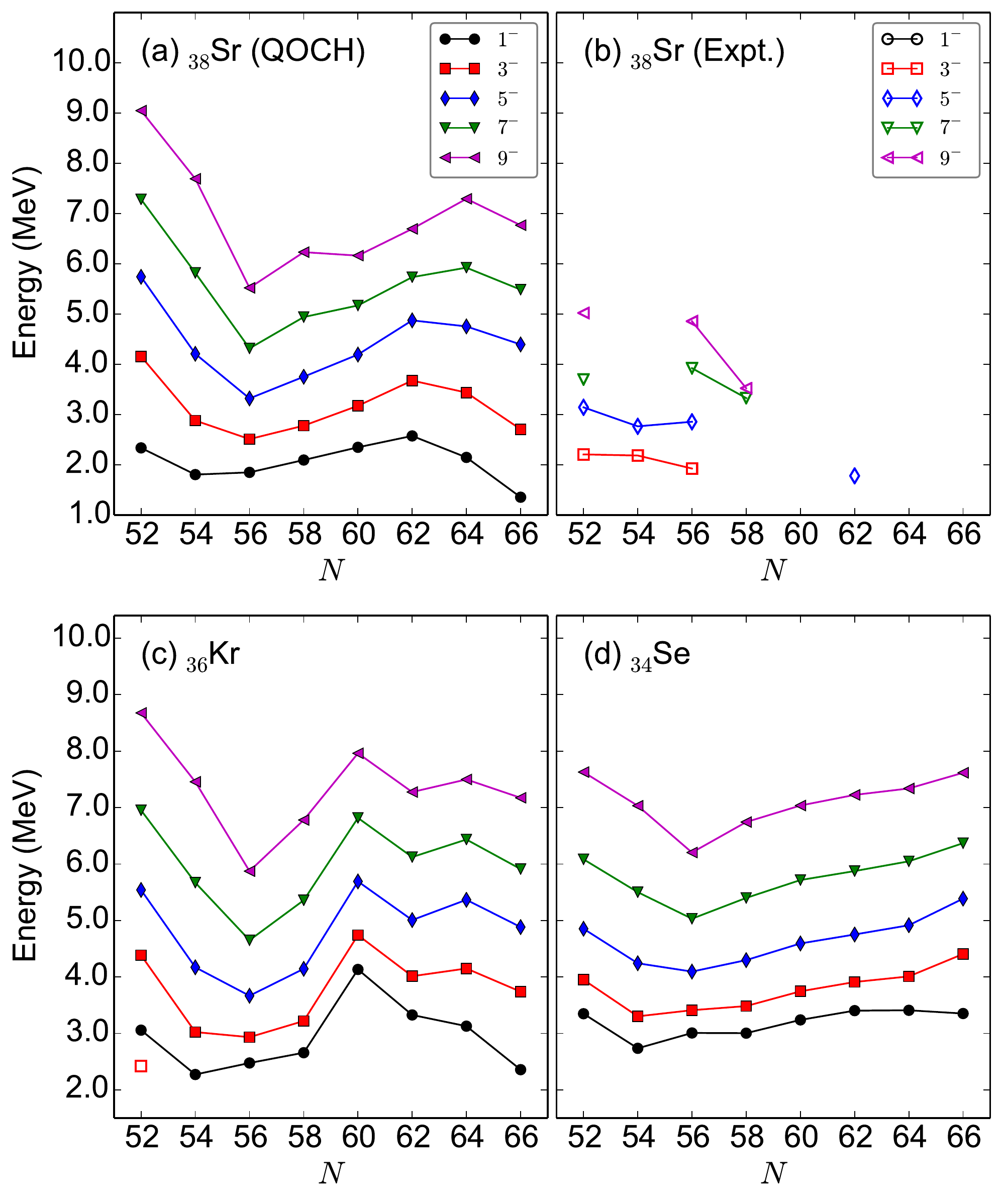} 
\caption{Same as in the caption to
 Fig.~\ref{fig:xe-level-pos} but for the negative-parity states in Se, Kr, and Sr isotopes.} 
\label{fig:level-kr-neg}
\end{center}
\end{figure}

%-----------------------------------------------------------------------
%
%	TRANSITION STRENGTH -- Xe-Ba-Ce
%
%-----------------------------------------------------------------------
\begin{figure*}[htb!]
\begin{center}
\includegraphics[width=0.7\linewidth]{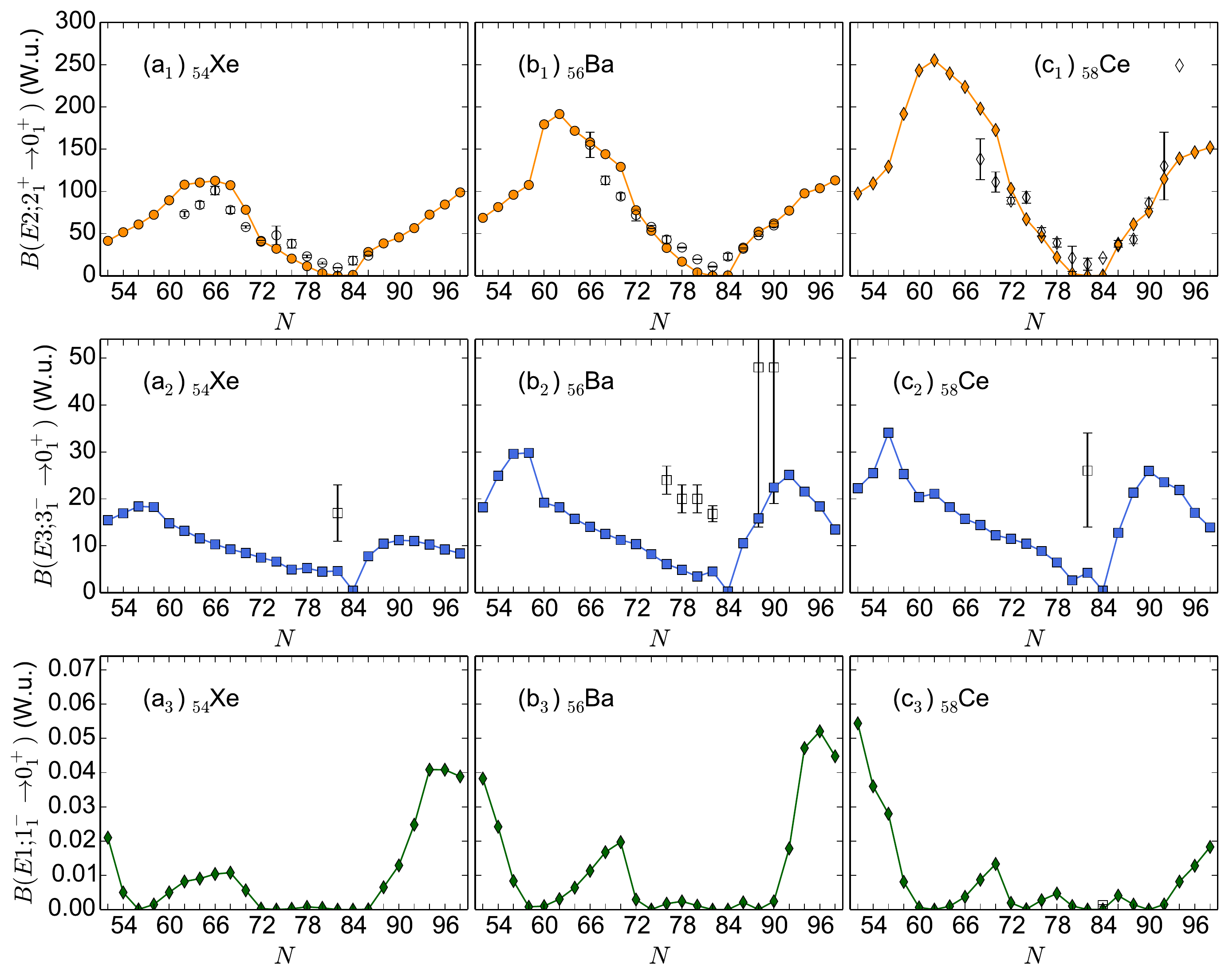} 
\caption{$B(\mathrm{E2};2^+_1\to 0^+_1)$ (upper),
 $B(\mathrm{E3};3^-_1\to 0^+_1)$ (middle), and
 $B(\mathrm{E1};1^-_1\to 0^+_1)$ (lower)
 reduced transition probabilities for the Xe, Ba, and Ce
 isotopes. Filled symbols connected by lines denote the QOCH
 results. Experimental values (open symbols) are taken from the ENSDF
 database \cite{data}.}
\label{fig:tr-xe}
\end{center}
\end{figure*}

%-----------------------------------------------------------------------
%
%	TRANSITION STRENGTH -- Kr
%
%-----------------------------------------------------------------------
%\begin{figure}[htb!]
%\begin{center}
%\includegraphics[width=\linewidth]{kr_tr.pdf} 
%\caption{Same as the caption to Fig.~\ref{fig:tr-xe}, but
% for the Se, Kr, and Sr isotopes.} 
%\label{fig:tr-kr}
%\end{center}
%\end{figure}
%-----------------------------------------------------------------------
%
%	SPECTRA -- N=56
%
%-----------------------------------------------------------------------
\begin{figure}[htb!]
\begin{center}
\includegraphics[width=\linewidth]{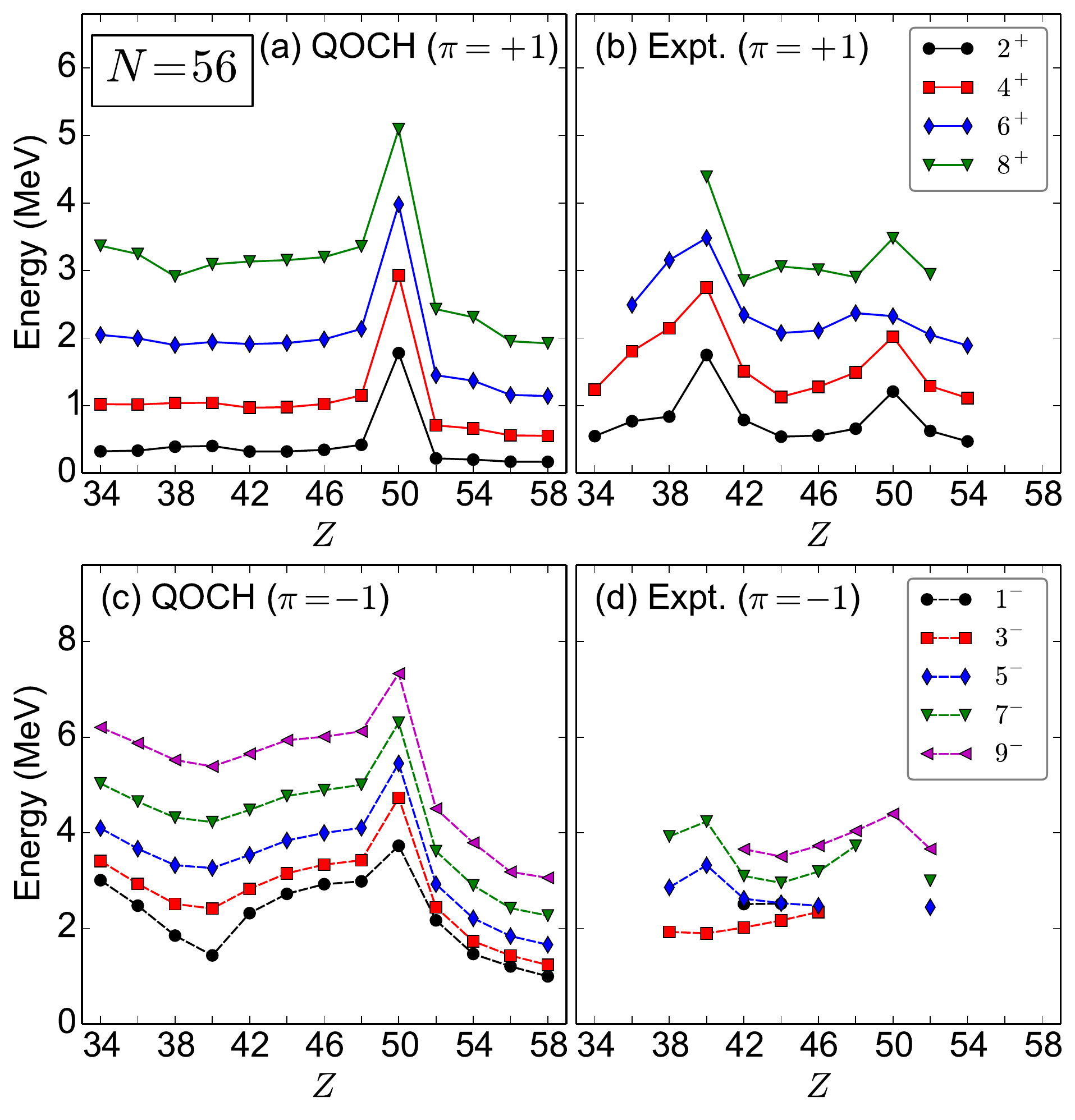} 
\caption{Excitation spectra of the low-lying positive-
 (upper row) and negative-parity (lower row) yrast 
 states of the $N=56$ isotones as functions of the proton number
 $Z$. The excitation spectra computed with the QOCH are plotted on the
 left-hand side of the figure, and are compared with the available
 experimental data \cite{data} on the right.}
\label{fig:level-n56}
\end{center}
\end{figure}

%-----------------------------------------------------------------------
%
%	B(E2), B(E3) -- N=56
%
%-----------------------------------------------------------------------
%\begin{figure}[htb!]
%\begin{center}
%\includegraphics[width=0.7\linewidth]{n56_tr.pdf} 
%\caption{Evolution of the $B(\mathrm{E2};2^+_1\to 0^+_1)$ and 
% $B(\mathrm{E3};3^-_1\to 0^+_1)$ transition rates of the $N=56$
% isotones as functions of the proton number $Z$. 
%Filled symbols represent the theoretical values and are connected by
% solid lines. 
%Open symbols in each panel stand for the experimental values taken from Ref.~\cite{data}.} 
%\label{fig:tr-n56}
%\end{center}
%\end{figure}

%-----------------------------------------------------------------------
%
%	average bet2 and bet3
%
%-----------------------------------------------------------------------
\begin{figure}[htb!]
\begin{center}
\includegraphics[width=\linewidth]{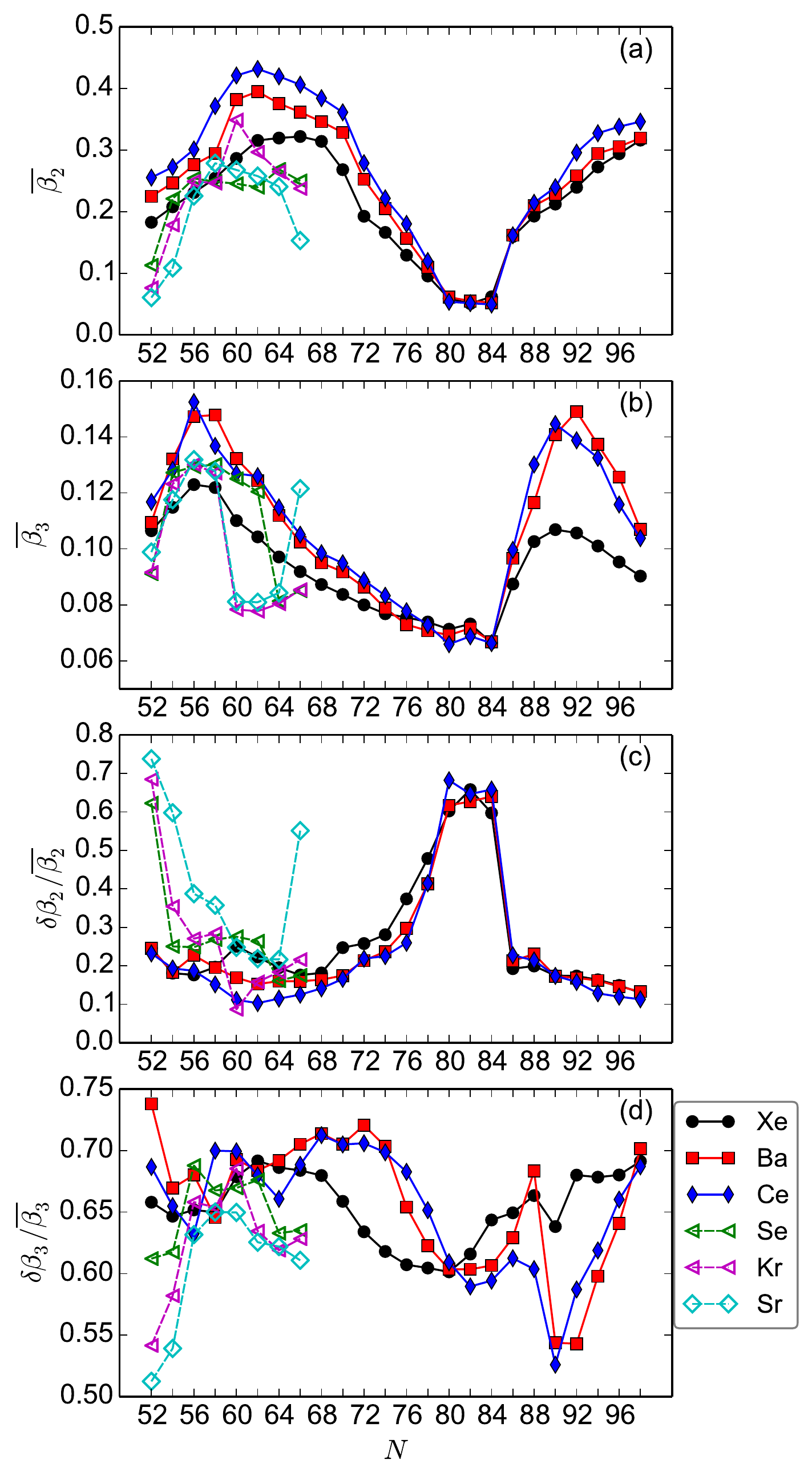} 
\caption{Average values of the $\beta_{2}$ (a) and
 $\beta_{3}$ (b) deformation parameters in the $0^{+}_{1}$ ground state,
 $\overline{\beta_{\lambda}}=\sqrt{\braket{\beta_{\lambda}^{2}}}$, and
 the fluctuations $\delta\beta_{2}/{\overline{\beta_{2}}}$ (c) and
 $\delta\beta_{3}/{\overline{\beta_{3}}}$ (d), for the Xe, Ba, Ce, Se,
 Kr, and Sr isotopes as functions of the
 neutron number $N$. The variance $\delta\beta_{\lambda}$ is defined as
 $\delta\beta_{\lambda}=\sqrt{\braket{\beta_{\lambda}^{4}}-\braket{\beta_{\lambda}^{2}}^{2}}/2\overline{\beta_{\lambda}}$. }
\label{fig:avb}
\end{center}
\end{figure}

\section{Spectroscopic results\label{sec:spectra}}

In the following we present QOCH results for the spectroscopic
properties relevant to quadrupole and octupole collective excitations. 
Note that the calculation also includes $N\approx Z\approx 56$ nuclei that are
close to the proton drip-line. For these nuclei only very limited experimental information
is available: the lightest known Xe, Ba, and Ce 
isotopes are $^{110}$Xe, $^{114}$Ba \cite{capponi2016}, and $^{118}$Ce. 
For completeness, and considering the signatures of octupole correlations on the 
corresponding $(\beta_{2},\beta_{3})$ PESs in
Fig.~\ref{fig:pes-prich}, we also discuss the spectroscopy of 
proton drip-line nuclei.

\subsection{Benchmark calculation:  $^{144}$Ba and  $^{112}$Xe}

As a test case, we consider the QOCH results for the low-energy
positive-parity ($\pi=+1$)
and negative-parity ($\pi=-1$) bands of $^{144}$Ba and $^{112}$Xe. 
These nuclei are specifically considered here as representative of the
regions close to the neutron octupole magic numbers $N\approx 88$ and
56. In particular, recent experiments performed at the Argonne National
Laboratory  \cite{bucher2016,bucher2017}, have indicated that 
the neutron-rich nucleus $^{144}$Ba and neighbouring
Ba isotopes are characterized by pronounced ground-state
octupole deformations. In Fig.~\ref{fig:level-baxe} we compare 
the lowest $\pi=+1$
and $\pi=-1$ QOCH bands of $^{144}$Ba and $^{112}$Xe to the available
data \cite{bucher2016,data}.
The predicted bands with both parities obtained from the present model
calculation are in a reasonable agreement with the experimental ones. 
The model also reproduces the data on E2 transitions in the $\pi=+1$
band of $^{144}$Ba, and predicts $B(E3)$ values for transitions between the
$\pi=+1$ and $\pi=-1$ bands. In $^{144}$Ba the calculated $B(E3;3^-_1\to
0^+_1)$ value of 16 W.u. is within the range of experimental uncertainty. 
%==================
% 5.3.2021
%
Previous GCM calculations for $^{144}$Ba based on the Gogny-D1S EDF
provide \cite{bernard2016,lica2018}
both positive- and negative-parity bands that are stretched as compared to the
experimental data, while the energy of the band-head state $1^{-}_{1}$ is
accurately reproduced. Moreover, the $B(E3; 3^{-}_{1}\to 0^{+}_{1})$
transition rates of $^{144}$Ba predicted by the recent GCM calculations
with both the Gogny-D1S EDF \cite{bernard2016} and the relativistic
functional PC-PK1 \cite{fu2018} are more or less similar to the one 
obtained in the present calculation. 
%==================
In the neutron-deficient $N\approx Z\approx 56$ region,  $^{112}$Xe is the
lightest nucleus for which experimental information is available. The
theoretical excitation spectrum is in qualitative agreement with the
data, though we note that the calculated $\pi=+1$ band appears to be
somewhat more compressed than the experimental one.

Figure~\ref{fig:cwf-baxe} plots the probability density distributions 
$\rho_{\alpha}^{I{\pi}}(\beta_2,\beta_3)$ (\ref{eq:rho}) of the lowest
energy positive ($0^+_1$) and 
negative-parity ($1^-_1$) states in the $(\beta_2,\beta_3)$-deformation space. 
One notices that, for both nuclei, the ground state
$0^+_1$ probability density is peaked at $\beta_2\approx\beta_{2,\mathrm{min}}$, where the
global minimum occurs on the PES, and $\beta_3\approx 0$. 
The collective wave functions for the $1^-_1$ state are, on the other
hand, concentrated at the same values of  $\beta_2$ as the corresponding ground states, but at 
finite values of the octupole deformation $\beta_3 \approx 0.1 - 0.15$.

\subsection{Low-energy excitation spectra}

The calculated excitation spectra for both even-spin positive- and 
odd-spin negative-parity yrast states are
shown in Figs.~\ref{fig:xe-level-pos} and \ref{fig:xe-level-neg},
respectively, 
for the Xe, Ba, and Ce isotopic chains in comparison with available data. 
The theoretical positive-parity states are in good agreement with 
experimental results (Fig.~\ref{fig:xe-level-pos}), with the exception 
of nuclei in the immediate vicinity of the neutron magic number $N=82$. For 
these nuclei the purely collective states of the QOCH cannot reproduce the 
empirical excitation spectrum on a quantitative level. 
%============================
% added on 3/3/2021
%============================
Within the the present calculation, the positive-parity bands in the
three isotopic chains are 
somewhat compressed as compared to the experimental values for
$52\leqslant N\leqslant 72$ and $N\geqslant 86$. This reflects the
fact that the SCMF potential surfaces for the corresponding nuclei
exhibit more pronounced $\beta_{2}$ deformations 
(Figs.~\ref{fig:pes-prich} and \ref{fig:pes-nrich}). 
In addition, there is a noticeable staggering
pattern of the predicted $6^{+}_{1}$ and $8^{+}_{1}$ excitation energies
around $N=56$, in particular, in the Xe isotopes
(Fig.~\ref{fig:pes-prich}(a$_{1}$)). 
As seen from the SCMF results in Fig.~\ref{fig:pes-prich}, the topology
of the PES varies rather rapidly from $^{110}$Xe to
$^{114}$Xe, that is, the degree of $\beta_{2}$ softness increases. As a
consequence, structures of the resultant 
positive-parity states could be significantly different between
neighbouring isotopes.

The results for the negative-parity states, shown in
Fig.~\ref{fig:xe-level-neg}, are more interesting. The calculated levels for 
each isotopic chain exhibit evident 
signatures of enhanced octupole collectivity, that is, a parabolic
behavior of excitation energies with neutron number, centered at around
$N\approx 56$ and $N\approx 88$. At these neutron numbers the levels
become lowest in energy. This is consistent with the observed trend of
the SCMF $(\beta_{2},\beta_{3})$ PESs in Figs.~\ref{fig:pes-prich} and
\ref{fig:pes-nrich}: in most of the nuclei around $N=56$ and 88 the
corresponding PESs display global minima at non-zero $\beta_{3}$.  
%====================
%  4.3.2021
%
A marked difference between the predicted and experimental $\pi=-1$
spectra is that the former increase rapidly as the neutron major shell
$N=82$ is approached, while the latter shows a flatter behaviour. 
Moreover, the quantitative agreement is not satisfactory for the $J=3^{-}$,
$5^{-}$, and $7^{-}$ excitation energies. 
As already mentioned, such discrepancies occur mainly because the
present QOCH framework only 
deals with the collective states. Particularly for the low-lying
negative-parity states in those nuclei near the magic
numbers, non-collective degrees of freedom come to play
a more relevant role. In such a case, phenomena like octupole
vibrations of spherical shape emerge, but they cannot be fully account
for within the present approach. 
%=====================

In Figs. ~\ref{fig:level-kr-pos} and \ref{fig:level-kr-neg} we
display the QOCH results for the excitation 
energies of the lowest positive- and negative-parity states in
neutron-rich Se, Kr and Sr isotopes, respectively. 
There are only few tentative assignments of negative-parity states in
neutron-rich $Z=34, 36, 38$ isotopes. Only few 
data for Kr isotopes are available. Also in this case one notices a kind of 
parabolic behavior centered at $N=56$, but much less pronounced than in the 
proton-rich Xe, Ba, Ce nuclei. Obviously in the latter case the $Z
\approx 56$ proton and 
$N=56$ numbers reinforce octupole correlations, and global minima at non-zero
$\beta_{3}$ are predicted. This does occur for the neutron-rich Se, Kr
and Sr isotopes, for which the corresponding PESs are at most soft in
the $\beta_{3}$ collective coordinates, as shown in Fig.~\ref
{fig:pes-kr}. 
In fact, it is well known that for these nuclei it is far more important to 
include the triaxial degree of freedom in order to describe the
excitation spectra at a quantitative level, and especially the shape
transition at $N=60$. A number of previous empirical and microscopic
studies have confirmed that the effects of triaxial deformations and
coexistence of different equilibrium shapes play an important role in
determining the low-energy nuclear structure around $N\approx 60$. 
The present version of the QOCH 
model is restricted to axially symmetric shapes and, therefore, the
calculated positive-parity spectra can only qualitatively reproduce the
empirical isotopic trend (Fig.~\ref{fig:level-kr-pos}). 
%=========================================
% added on 03/03/2021 for the comment 2. 
%=========================================
Similarly to the experimental trend, the predicted positive-parity levels
decrease as functions of $N$. 
The experimental data for the considered Se and Kr isotopes are
reasonably described, except for the nearly spherical nucleus $^{88}$Kr
($N=52$). 
Note that especially in Sr isotopes
(Fig.~\ref{fig:level-kr-pos}(a$_{2}$)) the experimental $\pi=+1$ spectra
show a marked peak at $N=56$. This points to the $N=56$ neutron subshell
closure due to filling of the $2d_{5/2}$ orbital. This systematics is
not observed in the 
predicted spectra,. This is expected from the fact that the corresponding
SCMF $\beta_{2}-\beta_{3}$ map for $^{94}$Sr is well deformed
(Fig.~\ref{fig:pes-kr}).

\subsection{Electromagnetic properties}

The results for the $B(E2;2^+_1\to 0^+_1)$, $B(E3;3^-_1\to 0^+_1)$, and
$B(E1;1^-_1\to 0^+_1)$ reduced transition probabilities along the Xe, Ba, and Ce
isotopic chains are shown in Fig.~\ref{fig:tr-xe}. We note a reasonable
agreement with the experimental $B(E2)$ values, especially considering
that bare charges are used in the calculation. 
Much less information is available on the $B(E3)$ values. 
What is interesting is that the theoretical $B(E3)$ values exhibit two
peaks, one at $N\approx 56$ and the other at $N\approx 88$. 
These neutron numbers, of course, correspond to the ones at which octupole
collectivity is most enhanced. Considering the
results on a more quantitative level, in each isotopic
chain the QOCH results for the $B(E3)$ rates systematically
underestimate the experimental values. The exceptions are
$^{144,146}$Ba, for which  
the experimental values are characterized by large uncertainties (see also Fig.~\ref{fig:level-baxe}). 
The reason why the QOCH cannot quantitatively reproduce the experimental 
 $B(E3)$ values is probably because of the fact that 
%the data are only available for 
for those nuclei that are close to the neutron magic number $N=82$, the
calculated energies of the $3^-$ state are also not in a particularly
good agreement with experimental results
(cf. Fig.~\ref{fig:xe-level-neg}). As shown in Fig.~\ref{fig:pes-nrich},
for those nuclei that are nearly spherical, there is no octupole
deformation or octupole softness at the SCMF level, so the collective
model is not expected to provide a very good description of E3 transition strength. 
There is no experimental information for the $E1$ transition
strengths. The systematics of the calculated $B(E1)$ values exhibits
certain peaks for particular nuclei, but they are not necessarily the
same as for the $B(E3)$ values. The $E1$ transitions are less
collective in nature compared to the $E2$ and $E3$ ones, hence the
collective model does not necessarily provide accurate predictions for 
the $B(E1)$ values.

\section{Systematics along the $N=56$ isotonic chains\label{sec:n56}}

We have also explored the systematics of excitation energies along the
$N=56$ isotones. The QOCH results for the low-energy positive-parity
and negative-parity spectra are shown in
Fig.~\ref{fig:level-n56}. The predicted positive-parity levels remain
almost constant with proton number $Z$, except for the $Z=50$ shell 
closure. The model qualitatively reproduces the corresponding 
experimental $\pi=+1$ spectra, with the exception of a pronounced 
proton-number dependence observed in the region 
$36\leqslant Z\leqslant 42$. The cusp in the experimental yrast states 
 indicates the $Z=40$ proton sub-shell closure, which is
not properly accounted for in the present calculation restricted to
axial symmetry. 
%=======================================
% added on 4.3.2021 for the comment 5
%
The failure in describing this experimental pattern could also be
attributed to the fact that both the employed energy density functional and
pairing property are not specifically adjusted to reproduce the $Z=40$
subshell closure. 
%=======================================
The computed $\pi=-1$ states become lowest in energy at $Z\approx 40$ for the
$Z<50$ region. Beyond the proton magic number $Z=50$, level spacing
between the negative-parity states are strongly reduced, and their energies display 
the parabolic trend characteristic of pronounced octupole correlations. The calculation also reproduces the empirical $B(E2)$ values, but underestimates the two known
$B(E3;3^-_1\to 0^+_1)$ at $Z=42$ and $Z=44$ by approximately a factor of 2. 

%In Fig.~\ref{fig:tr-n56}(a), the calculated $B(E2;2^+_1\to 0^+_1)$ values
%for the $N=56$ isotones are in a
%satisfactory agreement with the experiment. At $Z=38$ and 40, the
%present calculation considerably overestimate the experimental values,
%since the effect of $Z=40$ sub-shell does not enter but a stronger
%quadrupole collectivity emerges in the present calculation, as is seen
%in the PESs (see, Fig.~\ref{fig:pes-n56}). 
%In $50\leqslant Z\leqslant 58$, the $B(E2)$ values keep increasing, and are
%by a factor of two to three larger at than those below $N=50$. 
%A more or less similar $Z$-dependence is obtained for the $B(E3;3^-_1\to
%0^+_1)$ transition rates in Fig.~\ref{fig:tr-n56}(b), however the experimental values at $Z=42$ and 44 are considerably underestimated. 

\section{Signatures of octupole shape transitions\label{sec:qpt}}

As signatures of quadrupole and octupole shape transitions, we plot in
Fig.~\ref{fig:avb} the average values of the axial quadrupole
$\overline{\beta_2}$ (a) and $\overline{\beta_3}$ (b) deformation
parameters in the QOCH ground states $0^+_1$, and their fluctuations 
$\delta\beta_{2}/{\overline{\beta_{2}}}$ and
$\delta\beta_{3}/{\overline{\beta_{3}}}$, respectively, for the 
Ce, Ba, Xe, Sr, Kr, and Se isotopes, as functions of
the neutron number. Here the average $\overline{\beta_{\lambda}}$
($\lambda=2,3$) is defined as
$\overline{\beta_{\lambda}}=\sqrt{\braket{\beta_{\lambda}^2}}$, and
$\delta\beta_{\lambda}$ denotes the variance
$\delta\beta_{\lambda}=\sqrt{\braket{\beta_{\lambda}^4}-\braket{\beta_{\lambda}^{2}}^{2}}/2\overline{\beta_{\lambda}}$
\cite{li2010,srebrny2006}. 
In Fig.~\ref{fig:avb}(a), as expected from both the SCMF 
$(\beta_{2},\beta_{30})$ PESs and the calculated excitation spectra, 
the average deformation
$\overline{\beta_2}$ increases 
towards the middle of the major shell, as the quadrupole collectivity
becomes larger. 
The octupole deformation $\overline{\beta_3}$ exhibits a parabolic
behavior in two regions, centered at the neutron numbers $N=56$
and 88, at which it reaches maximum values larger than
$\overline{\beta_3}\approx 0.15$.  

In Fig.~\ref{fig:avb}(c), fluctuations of the $\beta_{2}$ deformation for
the Se, Kr, and Sr isotopes change abruptly from $N=58$
to 60. This reflects the
rapid structural evolution in these nuclei, most noticeably in Kr,
the relevant spectroscopic properties indicating phase-transitional
behavior at $N=60$. The
fluctuations of $\beta_{2}$ for the Xe, Ba, and Ce isotopes exhibit 
only a moderate change. This is consistent with the SCMF
results that the minima are more rigid in $\beta_{2}$. 

The fluctuation in octupole deformation $\beta_{3}$, depicted in
Fig.~\ref{fig:avb}(d), presents a measure for octupole
softness. Especially for the Sr and Kr nuclei near $N=54-56$, and
for the Ba and Ce nuclei from $N=88$ to 90, we observe a marked 
discontinuity characteristic of octupole shape-phase transitions. The 
isotopic dependence of the fluctuation in Fig.~\ref{fig:avb}(d)
correlates with the systematics of spectroscopic properties. 

%=================================
%   for the comment 4 (4/3/2021)
%
We note that in Fig.~\ref{fig:avb}(a) the average
$\overline{\beta_{2}}$ has a finite value $\approx 0.05$ for those 
Xe, Ba, and Ce nuclei near the magic number $N=82$. 
This is at variance with the SCMF result, in which the equilibrium minimum is found at
$(\beta_{2},\beta_{3})\approx (0,0)$ for the $N\approx 82$ nuclei. 
However, the average $\overline{\beta_{2}}$ is here obtained
by using the wave functions resulting from the diagonalization of the
QOCH and, therefore, does not necessarily coincide with the
equilibrium minimum in the SCMF calculation. Similar results were
obtained in a previous SCMF plus five-dimensional collective Hamiltonian
approach \cite{xiang2018}. We have also confirmed that the calculated
$0^{+}_{1}$ wave functions for the Ba isotopes with $N=80$, 82 and 
84 have the largest probability amplitudes  at $|\beta_{2}|\approx 0.05$. 
Finally, the large
$\beta_{2}$ fluctuations at the $N=82$ magic number are due to
vanishing values of $\overline{\beta_{2}}$ in the denominator in spherical
nuclei. 
%=================================

\section{Conclusions\label{sec:summary}}

Octupole collective excitations were analyzed using the fully microscopic
framework of nuclear density functional theory. Axially-symmetric
quadrupole-octupole constrained SCMF
calculations based on a choice of universal energy density functional
and pairing interaction were performed in three mass regions of the
nuclear chart in which
enhanced octupole correlations are empirically expected to occur: 
neutron-deficient nuclei with $Z\approx 56$ and $N\approx 56$,
neutron-rich nuclei with $Z\approx 56$ and $N\approx 88$, and the
neutron-rich nuclei with $Z\approx 34$ and $N\approx 56$. 
The resulting potential energy surfaces in the $(\beta_2,\beta_3)$ plane
indicate octupole-deformed equilibrium states at the SCMF level in
$^{112,114}$Ba and $^{114}$Ce on the neutron-deficient side, and in a
number of neutron-rich Ba and Ce nuclei around $N=88$. 

The SCMF calculations completely determine the
ingredients of the quadrupole-octupole collective Hamiltonian: the moment of
inertia, three mass parameters, 
and the collective potential. The diagonalization of the QOCH
subsequently yields 
the positive- and negative-parity excitation spectra and the electric
quadrupole, octupole, and 
dipole transition strengths that are relevant to the quadrupole and
octupole modes of 
collective excitations. 
%Considering the facts that the energy density functional is not
%specifically adjusted to the octupole modes and that the parameters of
%the collective Hamiltonian are completely determined by the SCMF
%calculations, 
The predicted excitation
spectra for both parities, and the $B(E2)$ and $B(E3)$ values of the
neutron-deficient and neutron-rich $Z\approx 56$ nuclei are in a
reasonable agreement with the experimental data. 
These quantities indicate a parabolic systematics
around the neutron numbers $N=56$ and 88, at which the SCMF
$(\beta_2,\beta_3)$ PESs exhibit pronounced octupole deformed minima. 
The calculated spectroscopic properties for the neutron-rich nuclei with
$Z\approx 34$ and $N\approx 56$ also indicate a signature of enhanced octupole
collectivity around $N\approx 56$, though not as distinct as in the case
of the $Z\approx 56$ isotopes.  
We have further explored spectroscopic properties
along the $N=56$ isotones, from $Z=34$ to 58. The relevant quantities,
i.e., negative-parity spectra and $B(E3)$ transitions, again point to
an enhancement of octupole correlations around $Z=34$ and 56. 
In general, octupole collectivity appears to be more enhanced in the
$N\approx 88$ region than around $N\approx 56$. The present
fully-microscopic spectroscopic calculation
has predicted several nuclei with stable octupole deformation in the
neutron-deficient $Z\approx 56$ nuclei, that have not been investigated so
far. The average $\beta_{2}$ and $\beta_{3}$ deformations calculated in
the QOCH ground states, as well as their fluctuations, exhibit 
signatures of quadrupole- and octupole shape phase transitions.

The current implementation of the QOCH method is restricted to
axially-symmetric shapes. Hence some discrepancies with the experimental
spectroscopic properties could be traced back to this 
limitation. In particular, the fact that the positive-parity states for
the neutron-rich $Z\approx 34$ and $N\approx 56$ nuclei have not been
reproduced quantitatively indicates that 
triaxial shape degrees of freedom need to be included as additional
collective coordinates. This 
requires the inclusion of several new terms in the collective Schr\"odinger
equation, but in practical applications such an extension would
be very complicated. Thus, a method that consists in
mapping the SCMF solutions onto the interacting-boson Hamiltonian
\cite{nomura2008} could be more feasible for the inclusion of triaxial degrees
of freedom. Work in this direction presents an interesting future study.

\begin{acknowledgments}
This work has been supported by the Tenure Track Pilot Programme of 
the Croatian Science Foundation and the 
\'Ecole Polytechnique F\'ed\'erale de Lausanne, and 
the Project TTP-2018-07-3554 Exotic Nuclear Structure and Dynamics, 
with funds of the Croatian-Swiss Research Programme. 
It has also been supported in part by the QuantiXLie Centre of Excellence, a project co-financed by the Croatian Government and European Union through the European Regional Development Fund - the Competitiveness and Cohesion Operational Programme (KK.01.1.1.01.0004).
\end{acknowledgments}

\bibliography{refs}

\end{document}